\documentclass[a4paper,twocolumn]{article} 

\usepackage{times}  
\usepackage{url}  
\usepackage{graphicx}  
\usepackage{subcaption}
\usepackage{amsmath}

\usepackage{color}
\usepackage{amsthm}
\usepackage{booktabs, tabularx} 
\usepackage{diagbox} 
\usepackage{authblk}
\usepackage{hyperref}

\newcommand{\learningphase}{early phase}
\newcommand{\exploitationphase}{late phase}
\newcommand{\rejectedrqone}{RQ1}
\newcommand{\rejectedrqtwo}{RQ2}
\newcommand{\MD}{MD}
\newcommand{\SU}{SU}

\newcommand{\negative}{\emph{Negative}}
\newcommand{\positive}{\emph{Positive}}
\newcommand{\irrelevant}{\emph{Irrelevant}}
\newcommand{\relevant}{\emph{Relevant}}
\newcommand{\factual}{\emph{Factual}}
\newcommand{\nonfactual}{\emph{Non-factual}}
\newcommand{\ambiguous}{difficult}
\newcommand{\nonambiguous}{easy}
\newcommand{\prednd}{PredictorE}
\newcommand{\predd}{PredictorD}

\begin{document}
%
\title{How Does Tweet Difficulty Affect Labeling Performance of Annotators?}
\date{}
\author[1]{Stefan R{\"a}biger\thanks{stefan@sabanciuniv.edu}}
\author[1]{Y\"{u}cel Sayg{\i}n}
\author[2]{Myra Spiliopoulou}
\affil[1]{Sabanc{\i} University (Istanbul, Turkey)}
\affil[2]{Otto-von-Guericke University (Magdeburg, Germany)}

\renewcommand\Authands{ and }
\maketitle
\begin{abstract}
Crowdsourcing is a popular means to obtain labeled data at moderate costs, for example for tweets, which can then be used in text mining tasks. To alleviate the problem of low-quality labels in this context, multiple human factors have 
been analyzed to identify and deal with workers who provide such labels. However, one aspect that has been rarely considered is the inherent difficulty of tweets to be labeled and how this affects the reliability of the labels that annotators assign to such tweets. Therefore, we investigate in this preliminary study this connection using a hierarchical sentiment labeling task on Twitter. We find that there is indeed a relationship between both factors, assuming that annotators have labeled some tweets before: labels assigned to \nonambiguous{} tweets are more reliable than those assigned to \ambiguous{} tweets. Therefore, training predictors on \nonambiguous{} tweets enhances the performance by up to 6\% in our experiment. This implies potential improvements for active learning techniques and crowdsourcing.
\end{abstract}
\section{Introduction}

Studies in crowdsourcing have found that labels assigned by workers\footnote{We use "worker" and "annotator" interchangeably in this work, where the former term is more suitable for a crowdsourcing environment, while the latter one is preferred in more general contexts.} to documents become
more reliable towards the end of a worker's labeling session \cite{gadiraju2015training,maddalena2016crowdsourcing,rabiger2018how}.
Similarly, the time needed to assign labels to documents drops rapidly in a worker's \learningphase{} until it converges to a roughly constant level in the \exploitationphase. Since annotation times are typically associated with labeling costs, shorter annotation times are preferred. Thus, when experimenters want to recruit workers on a crowdsourcing platform who are likely to assign high-quality labels, suitable workers should (a) have completed similar tasks before and (b) have reached the state where labeling
costs are approximately constant to keep the time needed for task completion short.

In practice, however, we suspect that this strategy could be affected by the inherent
difficulty of the documents to be labeled since some documents are more \ambiguous{} to label than others. Therefore, we expect that labels assigned to \ambiguous{} documents will be less reliable. Using these \ambiguous{} documents for training could affect the performance of the resulting predictors adversely. In contrast, if the reliability of the labels in the training set is high, resulting predictors could improve their performance. 
Thus, we assume that label reliability can be inferred from measuring the performance of predictors: given the performances of two predictors, we assume that the one achieving better performance was trained on documents with more reliable labels.

If this idea holds, we imagine to build in the future a difficulty predictor that estimates the difficulty level of documents in a preprocessing step to separate \ambiguous{} from \nonambiguous{} documents. For example, in crowdsourcing the difficulty could be trained on a small seed set and then estimate the difficulty of the remaining documents. Only \nonambiguous{} documents would then be retained. This could potentially help avoid wasting human labor and budget on \ambiguous{} documents which should not be annotated at all. 
Similarly, such a difficulty predictor would also be a helpful means as a preprocessing step in active (machine) learning \cite{settles2012active}.
Whenever a label for a document is to be requested in active learning, it is expected that the human oracle (annotator) provides a reliable label. If a document to be labeled is \nonambiguous, the label assigned by the annotator will be more likely reliable, but if the document is \ambiguous, then there might be no suitable label available due to the difficulty of the document. Hence, applying the difficulty predictor in advance would allow to invoke active learning strategies only for \nonambiguous{} documents.


The concept of "early" and "late" annotation phase is inspired by the observation that annotators need some time to learn how to annotate \cite{zhu2010analysis,settles2008active,rabiger2018how}. The time, translated here to the number of documents one sees, depends on the annotator. We roughly split the annotation process into an early phase encompassing the first $n$ documents and a late phase comprising the next $n$ documents. (In our experiments, some annotators labeled more than $2n$ documents, but we ignore these documents to avoid the effects of fatigue.)
We define "document difficulty" informally as the set of factors that determine to what extend workers are hesitant in choosing among the available labels for a document. These factors may be features of the document, e.g. words in the document, but may also be in the eye of the beholder, e.g. affected by the workers' perception of and attitude towards the subject matter.
Since we cannot fix the factors making a document \ambiguous{} as solely inherent to the document, we rather rely on difficulty indicators, which are labeling cost, worker disagreement \cite{sameki2016dynamic} and predictor certainty \cite{alonso2015predicting}.
Then, we propose predictors of annotator performance and study how phase and tweet difficulty influence the expected label of a document. 

Since modeling the difficulty of tweets has been rarely the subject of investigation, we use the dataset from \cite{rabiger2018how}.
Another advantage of this dataset is that sentiment analysis is known to be subjective and therefore sufficiently \ambiguous. This difficulty is also perceived by crowd workers \cite{gadiraju2014taxonomy}, which allows us studying the interplay between tweet difficulty and the label reliability in annotators' early/late annotation phase.  
To the best of our knowledge, this problem has not been analyzed before. 
Specifically, we address the following research questions in this report:
\begin{itemize}
    \item \textbf{\rejectedrqone}. How does document difficulty in the training set affect the performance of resulting predictors in the \learningphase{} and in the \exploitationphase?
    \item \textbf{\rejectedrqtwo}. Are these effects from \rejectedrqone{} meaningful?
\end{itemize}

Our analysis should be regarded as a preliminary study because the dataset is relatively small. However, if there is a connection between label reliability and document difficulty, in the next step real crowdsourcing experiments can be performed. This is a common approach in crowdsourcing, e.g. \cite{yang2013freeloc,alonso2009can,salomoni2015crowdsourcing}, for multiple reasons. For one, budget may be saved if proposed methods turn out not to work. Another reason is that one might want to run an experiment first in a controlled environment, as done in \cite{rabiger2018how}, to avoid external influence factors which cannot be ensured in crowdsourcing.


\section{Related Work}
The most relevant literature for our work addresses how document difficulty, and in particular tweet difficulty, is modeled in crowdsourcing and similar environments.

Martinez et al. utilize a predictor's certainty to approximate the difficulty of a document \cite{martinez2013document}. The underlying assumption is that a predictor is less certain about predicting labels for \ambiguous{} documents. We employ the same idea in this work to derive tweet difficulty heuristically.
Label difficulty has also been acknowledged and researched in the context of active learning \cite{culotta2005reducing} and crowdsourcing \cite{gan2017incentivize}. However, Gan et al. \cite{gan2017incentivize} focus on modeling the difficulty of labeling tasks in crowdsourcing instead of single documents. Paukkeri et al. \cite{paukkeri2013assessing} propose a method to estimate a document's subjective difficulty for each user separately based on comparing a document's terms with the known vocabulary of an individual. 
Sameki et al. model tweet difficulty in the context of crowdsourcing \cite{sameki2016dynamic} where they devise a system that minimizes the labeling costs for micro-tasks by allocating more budget to difficult tweets and less to easy ones. The authors argue that more sentiment makes a tweet more difficult to understand. Hence, they formulate the problem of estimating tweet difficulty as a task of distinguishing sarcastic from non-sarcastic tweets. One of the factors that they utilize is annotator disagreement - if more individuals agree on a label, it is considered easier.
An approach that is related to this idea in spirit exists for estimating the difficulty of queries \cite{carmel2006makes}: topic difficulty is approximated by analyzing the performances of existing systems - a lower performance indicates more difficult topics.
In our work, we also harness annotator disagreement to approximate tweet difficulty - lower annotator disagreement is associated with easier tweets. While our work bears similarities with \cite{sameki2016dynamic}, the objectives differ: we are explicitly interested in analyzing how tweet difficulty affects the reliability of tweets that annotators assign, while Sameki et al. employ tweet difficulty as a feature to predict the number of annotators who should label the tweet. Furthermore, we combine worker disagreement with two more factors to model tweet difficulty.
Another related approach is described in \cite{whitehill2009whose} where the authors propose a probabilistic method that takes image difficulty and crowd worker expertise into account to derive a ground truth -- the authors show that this idea is more accurate than majority voting. However, they do not consider that workers learn during a labeling task. In addition, we focus on analyzing how the performance of predictors is affected by tweet difficulty.

Although we are investigating tweet difficulty in crowdsourcing, we do not analyze any online crowdsourcing activity \cite{bontcheva2017crowdsourcing,yang2016enhancing} on tweets, because we first need to know how annotators behave in a fully controlled experiment, before we include the uncertainty associated with worker diversity/background knowledge and engagement/disinterest. Similarly, in this work we do not discuss human factors, e.g. how worker expertise affects label reliability \cite{kazai2013analysis} because we performed an experiment in a controlled environment with volunteers which we consider faithful. Likewise, the annotators share a similar background in that they are computer science students.


Despite tweets being text documents, we do not use any of the proposed methods, e.g. \cite{goncalves2017task}, to model difficulty. This is because tweets are too short to extract meaningful grammatical features and sometimes they even do not contain any well-formed sentences at all. Therefore, we model tweet difficulty using the abovementioned heuristics from the crowdsourcing context which correlate intuitively with tweet difficulty. 

\section{Our Approach}
We first describe briefly the dataset we use for performing our experiment. This is followed by addressing the different steps involved in designing our experiment for the analysis of the research questions.  

\subsection{Description of Dataset}
\begin{figure}
\centering
\includegraphics[width=\columnwidth]{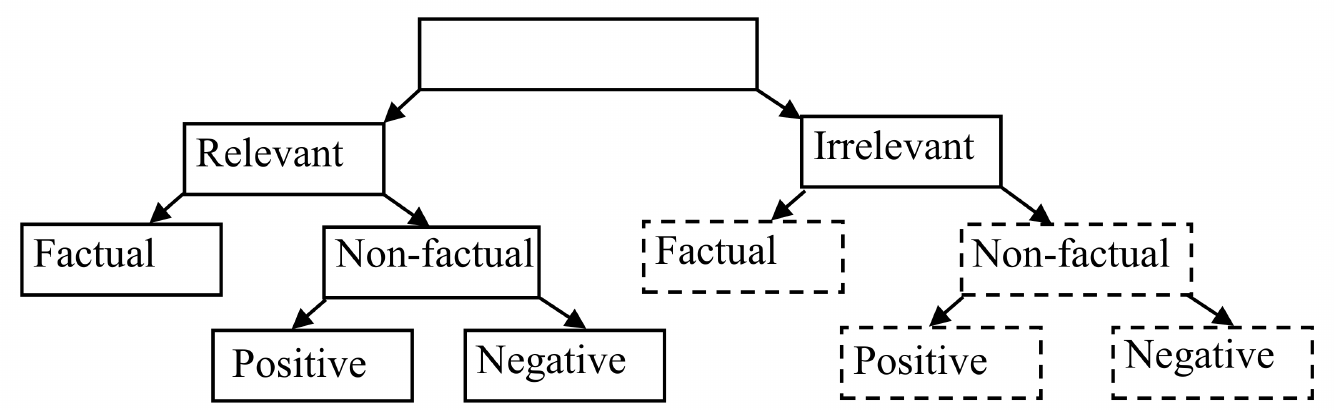}
\caption{Hierarchical labeling scheme. Labels with dashed lines were removed from the dataset. Each hierarchy level corresponds to a label set: the first set is Relevant/Irrelevant, the second one is Factual/Non-factual, and the third one is Positive/Negative.}
\label{fig:labels}
\end{figure}

In our analysis we use the dataset from \cite{rabiger2018how}, which contains 500 tweets labeled hierarchically in terms of sentiment in two geographically different regions, Magdeburg (\MD) and Sabanc{\i} (\SU). Conducting our experiment for both regions separately reduces the chances of our results being coincidence or biased by location-specific factors.
The collected tweets address the first US presidential debate between Hillary Clinton and Donald Trump in 2016. One sample tweet is shown below:
\begin{verbatim}
Did trump just say there needs to be 
law and order immediately after 
saying that he feels justified not 
paying his workers??  #Debates
\end{verbatim}

The hierarchical labeling scheme for this dataset is depicted in Fig.~\ref{fig:labels} and comprises three levels. On the first level, a tweet is either \relevant{} or \irrelevant{} with respect to the topic and on the second level either \factual{} (= neutral) or \nonfactual. If a tweet is considered \nonfactual{}, it is either \positive{} or \negative{} on the third level. For \irrelevant{} tweets any additional labels (e.g. \factual) and their corresponding metadata (annotation times) are ignored as we are only interested in the sentiment of \relevant{} tweets. Note that this labeling scheme enforces annotators to assign no sentiment to neutral tweets.


In total, a similar number of students participated in the annotation process: 19 in \MD{} and 25 in \SU{} as depicted in Table~\ref{tab:anno_dist}.

\begin{table}
    \centering
    \begin{tabular}{ccccc} 
      \toprule
      Group & \multicolumn{3}{c}{Tweet set size} & Total \\
      & S & M  & L & \\
      \midrule
      \MD & 10 & 8 & 1 & 19 \\
      \SU & 13 & 9 & 3 & 25 \\
      \bottomrule
    \end{tabular}
    \caption{Annotator distribution and total number of labeled tweets per group. S is 50 tweets, M is 150 tweets, and L is 500 tweets.}
\label{tab:anno_dist}
\end{table}

\subsection{Measuring Tweet Similarity}
Since we employ a kNN predictor\footnote{We opted for kNN as it considers neighborhoods and we believe that the type of difficulty we investigate is a local phenomenon ("Are similar tweets \ambiguous{} or \nonambiguous{} to label?"), so we do not want to use an SVM or similar predictors as they learn globally optimal models ("Is the tweet \nonambiguous{} or \ambiguous{} to label?")} in our experiment, we must compute the similarity between any two tweets $t1$ and $t2$. As a result of tweets exhibiting different lengths, we normalize this similarity by the longer tweet to avoid any influence of the text length on the similarity. Therefore, this normalized similarity yields values between zero (tweet texts are disjoint) and one (identical tweets). We refer to this normalized similarity as $Nsim$ and it is computed between $t1$ and $t2$ as:
\begin{equation}
    \label{eq:dist}
    Nsim(t1, t2) = sim(w1,w2)/ max(|w1|,|w2|) 
\end{equation}

where $w1$ and $w2$ represent the words in the tweets $t1$ and $t2$ and $sim(w1,w2)$ computes the number of shared words between $t1$ and $t2$ according to a similarity metric. In this preliminary study, we utilize as metric the same three metrics that were used in \cite{rabiger2018how}, namely longest common subsequence, longest common substring, and edit distance .

These three metrics are typically defined on character-level, i.e. they compute the similarity between two single words by comparing these words character by character. Since we deal with tweets containing multiple words, we apply them on word-level.
For example, edit distance between two strings usually counts how many characters in one string need to be changed to transform it into the other one. However, we count how many words in $t1$ must be replaced s.t. it results in $t2$. Longest common subsequence counts how many characters in both words are in the same relative, but not necessarily contiguous, order.  Extending this to tweets means we now count the words in $t1$ and $t2$ that are in the same relative, but not necessarily contiguous, order. Similarly, longest common substring counts how many contiguous characters both words share. That means in our case we count the number of words that are contiguously shared among $t1$ and $t2$.

For $Nsim$ to yield values between zero and one, the term $sim(w1,w2)$ needs to be inversed when using edit distance because large values indicate that $t1$ and $t2$ are different as opposed to being similar. Thus, when using edit distance, we use $1-sim(w1,w2)$  instead of $sim(w1,w2)$ in the numerator of Equation~\ref{eq:dist}.

\subsection{Modeling Annotation Difficulty}
\label{sec:difficulty}
Since there is no ground truth for tweet difficulty available, we approximate the difficulty of a tweet $t$ by computing its difficulty score $DS$. $DS(t)$ combines three heuristics, namely worker agreement ($A$) \cite{sameki2016dynamic}, predictor certainty ($C$) \cite{martinez2013document}, and labeling cost ($L$): 
\begin{equation}\label{eq:ds}
DS(t)=A+C+L         
\end{equation}
where $A,C,L\in[0,1]$. We define higher difficulty scores in this equation to correspond to easier tweets. 

The labeling agreement $A$ measures the extent to which annotators agree on a label, where higher values indicate easier tweets. To compute $A$ for $t$, we devise a scoring function $A(t)$ yielding values between 0 (no agreement) and 1 (perfect agreement).  Furthermore, the worker agreement of each hierarchy level must contribute to $A$. Specifically, we use majority voting to assign a label to each hierarchy level. A level should contribute more to $A$ if more workers agreed on the label. Since lower hierarchy levels might have been labeled by less workers than the first level (namely if workers deemed a tweet \irrelevant{} or \factual), higher levels tend to contribute more to $A$. This reasoning is reflected in the following equation:
\begin{equation}
\label{eq:1}
A(t)=\sum_{i\in~Levels}{\frac{|annotators_{maj}|}{|annotators_i|}*\frac{|annotators_{maj}|}{total_{maj}}}
\end{equation}
where $annotators_{maj}$ are the annotators who assigned the majority label on hierarchy level $i$, $annotators_i$ are the annotators who labeled $t$ on level $i$, $total_{maj}$ is the total number of annotators across all hierarchy levels that assigned majority labels, and $Levels$ is the set of hierarchy levels in the labeling scheme, in our case $Levels=\{1,2,3\}$.
The first term in the product describes the fraction of annotators who agreed on level $i$ on the majority label, while the second expression accounts for the overall contribution of level $i$ to the overall agreement. Whenever there is a tie on level $i$ regarding the majority label,  $total_{maj}$ is incremented by one. This lowers the contribution of levels that have no ties to the overall labeling agreement, which generally leads to lower agreement ratings for tweets with ties.

The following two examples illustrate how Equation~\ref{eq:1} approximates annotator agreement. First, suppose that four annotators labeled tweet $t1$ and assigned the labels:
\begin{itemize}
    \item First hierarchy level: \relevant, \relevant, \relevant, \relevant 
    \item Second hierarchy level: \factual, \nonfactual, \nonfactual, \nonfactual
    \item Third hierarchy level: -, \negative, \negative, \positive
\end{itemize}

Therefore, the majority labels for $t1$ are \relevant{}, \nonfactual, and \negative, leading to $A(t1)= 4/4*4/9 + 3/4*3/9 + 2/2*2/9=0.92$. In total, nine workers assigned the majority labels (four on the first level, three on the second level, two on the third level), so $total_{maj}=9$. In the second example, suppose there was a tie on the second level of $t1$, i.e. 
\begin{itemize}
    \item First hierarchy level: \relevant, \relevant, \relevant, \relevant 
    \item Second hierarchy level: \factual, \nonfactual, \nonfactual, \factual
    \item Third hierarchy level: -, \negative, \negative, -
\end{itemize}

This time there are two possibilities for the majority labels: either \relevant{} and \factual{} or \relevant{}, \nonfactual, and \negative{}. In this case the majority labels would be chosen randomly. Regardless of that outcome, the resulting worker disagreement score would be now $A(t1) = 4/4*4/9 + 2/4*2/9 + 2/2*2/9=0.78$. Note that in this case $total_{maj}=9$ instead of $total_{maj}=8$ because exactly one tie occurred on the second hierarchy level, leading to a lower agreement score than in the first example.

A higher predictor certainty $C$ for a tweet indicates easier tweets. To compute it, we build a kNN predictor for each annotator separately since sentiment is subjective. The predictor is trained on 40\% of an annotator's labeled tweets and the longest common substring\footnote{We obtained similar results when choosing edit distance or longest common subsequence.} is used to compute the similarity between any pair of tweets. Since kNN does not naturally provide a certainty for the predicted label $j$ of tweet $t$, we approximate it as follows:
\begin{equation}
\label{eq:certainty}
certainty_j(t) = \frac{n_j + s}{k+c}
\end{equation}
where $n_j$ is the number of the $k$ neighbors that share label $j$, $s$ being a smoothing factor to avoid zero probabilities, and $c$ being the number of possible classes that exist on a certain hierarchy level. In our experiment we set $s=1$.
We store for each tweet of a worker's test set (60\% of the labeled tweets) the certainty $C$ of the predicted labels.
Repeating this process for all workers yields a list of predictions per tweet on each hierarchy level. 
To obtain a single certainty per tweet, we first average the certainties (of the different workers who labeled the tweet) per level and from these certainties we pick the maximum certainty per level, i.e. this process yields three values. Each of these three certainties corresponds to the predicted majority label on the respective hierarchy level. Averaging these three values yields $C(t)$. This procedure is reflected in the following equation: 
\begin{equation}\label{eq:pc}
C(t) = \frac{1}{3}\sum_{i\in{}Levels}\max_{j\in{}Labeled} \frac{ \sum_{k\in{}Workers} certainty_j(t)}{|Workers|}
\end{equation}
where $Labeled$ is the set of predicted labels for $t$ on hierarchy level $i$, $Workers$ is the set of annotators who labeled $t$ in their test sets, and $Levels$ is the set of hierarchy levels in the labeling scheme, in our case $Levels=\{1,2,3\}$.
Note that in this procedure we are not accessing the sentiment labels which kNN predicts for a tweet. Instead, we only use the predictor certainties of the sentiment labels that kNN assigned to the tweets. Therefore, we are not leaking any  information about the actual sentiment labels to the sentiment predictors that are built in the experiment. Table~\ref{tab:aggregating_certainties} illustrates how $C(t1)$ is obtained for $t1$. In this case two annotators have $t1$ in their test set, hence we have four predictor certainties (two predicted labels per worker) per level. For example, kNN is 80\% certain, according to Equation~\ref{eq:certainty}, that worker 1 (first row, first column) would assign \relevant{} to $t1$ on the first hierarchy level. In contrast, kNN is only 20\% certain for her to assign \irrelevant{}. The certainties are averaged per label and per level (row 3), e.g. the average certainty of kNN to assign \relevant{} on the first hierarchy level is $(80\% +70\%)/2=75\%$, while it is $(20\%+30\%)/2=25\%$ for \irrelevant. Averaging these three remaining certainties results in $C(t1)=68\%$.

\begin{table*}[t]
\centering
  \begin{tabular}{cccc}
    \toprule
    & First level & Second level & Third level \\
    \midrule
    Annotator 1 & ($R$, .8), ($IR$, .2) & ($F$, .4), ($NF$, .6) & ($P$, .3), ($N$, .7) \\
    Annotator 2 & ($R$, .7), ($IR$, .3) & ($F$, .2), ($NF$, .8) & ($F$, .3), ($NF$, .7) \\
    Avg. certainty & ($R$, .75), ($IR$, .25) & ($P$, .5), ($N$, .5) & ($P$, .4), ($N$, .6) \\
    Maximum certainty & .75 & .7 & .6  \\
    $P(t1)$ & & $(.75+.7+.6)/3 = .68$ & \\
  \bottomrule
\end{tabular}
\caption{Example how Equation~\ref{eq:pc} aggregates the predicted certainties for tweet $t1$. The columns represent the hierarchy levels in the labeling task. We use the following acronyms to represent the predicted sentiment labels: $R$: \relevant, $IR$: \irrelevant, $F$: \factual, $NF$: \nonfactual, $P$: \positive, $N$: \negative. Suppose two annotators labeled $t1$ in their test sets and kNN predicted for each worker a tuple of (sentiment label, certainty) according to Equation~\ref{eq:certainty} per hierarchy level. "Avg. certainty" averages the predicted certainties per label per hierarchy level. "Maximum certainty" shows which certainty would be kept according to Equation~\ref{eq:pc} and the last row shows the final result of the computation, thus $C(t1)=0.68$ in this case.}
\label{tab:aggregating_certainties}
\end{table*}

The labeling cost $L$ for tweet $t$ corresponds to $t$'s median annotation time. The higher it is, the more \ambiguous{} it is to label $t$. However, since high values of $DS(t)$ are associated with \nonambiguous{} tweets, $L$ must be inverted. We choose as labeling cost for $t$ the median annotation time across all annotators who labeled it. The median is more appropriate than the average in our case due to its robustness toward outliers because some annotators had a few random spikes in their annotation times. After normalizing the labeling cost, the following equation follows:
\begin{equation}
L(t)=1-\frac{cost_t-cost_{min}}{cost_{max}-cost_{min}}
\label{eq:cost}
\end{equation}
where $cost_t$ is the median labeling cost of tweet $t$, $cost_{min}$ ($cost_{max}$) is the lowest (highest) median labeling cost across all tweets.


After computing $DS$ for each tweet, we apply k-means with $k=2$ to cluster the difficulty scores. Each tweet is now assigned a difficulty label, \nonambiguous{} or \ambiguous,  
according to its cluster membership.

\subsection{Design of the Simulation Experiment}
\label{sec:simulation_experiment}
By training predictors we want to answer \rejectedrqone, i.e. if \ambiguous{} tweets affect label reliability in the \learningphase{} and in the \exploitationphase. The goal is to predict the hierarchical sentiment labels (\relevant, \irrelevant, \factual, \nonfactual, \positive, \negative). We measure predictor performance in terms of hierarchical F1-score, which is recommended by Kiritchenko et al. for hierarchical labeling tasks \cite{kiritchenko2006learning}.  
Specifically, we analyze the effect of the following independent variables on predictor performance: 
\begin{itemize}
    \item \textit{difficulty}: \ambiguous{} or \nonambiguous{} tweets
    \item \textit{phase}: \learningphase{} or \exploitationphase (cf. Section~\ref{sec:lp_ep})
    \item \textit{training set size}: number of tweets in the training set 
    \item \textit{neighbors}: number of nearest neighbors in kNN
    \item \textit{institution}: either \MD{} or \SU
\end{itemize}
We expect meaningful patterns observed in this simulation to hold while varying the abovementioned variables. Otherwise the patterns might be due to chance. For example, if one predictor outperforms another one, this result should hold even if the size of the training set changes.

The core assumption in this simulation experiment is that the reliability of labels can be inferred from measuring the performance of trained predictors: if predictors achieve higher F1-scores, the sentiment labels in their training sets are considered more reliable. In other words, we use F1-score as a proxy for the reliability of labels.  
Therefore, we train two predictors per crowd worker, \prednd{} trained only on \nonambiguous{} tweets and \predd{} which is trained solely on \ambiguous{} tweets. We fix all of the abovementioned variables, so that only the variable \textit{difficulty} of the training set differs between both predictors. This allows us to draw conclusions about the effect of tweet difficulty on label reliability.

\subsection{Early \& Late Phase in Annotator Behavior}
\label{sec:lp_ep}
For the experiment, our dependent variable -- predictor performance -- is affected by two parameters: the number of tweets used in the training set and tweet difficulty. That means we plot a curve of the predictor performances once for \ambiguous{} and once for \nonambiguous{} tweets while varying the number of tweets in the training set.
However, annotators undergo a \learningphase{} \cite{maddalena2016crowdsourcing,zhu2010analysis, rabiger2018how}, i.e. a drop in annotation times occurs in the beginning of an annotation session. Thus, the phase -- either \learningphase{} or \exploitationphase -- is also an independent variable that we need to control for in our experiment. Therefore we perform the experiment once for  the \learningphase{} and once for the \exploitationphase{} because within these phases the annotation times can be considered similar.

Originally, annotators labeled either S, M, or L tweets in the used dataset according to their annotator group and it was found that the length of the \learningphase{} differs across the annotator groups \cite{rabiger2018how}. To avoid having to control for this variable as well, i.e. repeating the experiment with the two phases once for each annotator group, we fix the length of the \learningphase{} across all three annotator groups. When aggregating all annotation times per institution, either \MD{} or \SU, we obtain for the length of the \learningphase{} approximately 25 tweets, i.e. the first 25 labeled tweets of each annotator are used for their \learningphase{} and their next 25 labeled tweets are utilized for their \exploitationphase{} to have a balanced experimental setup. Therefore, we use in total the first 50 labeled tweets of each annotator in both institutions. Any other labeled tweets are discarded.  Another reason for not using more tweets for the \exploitationphase{} is to avoid uncontrollable side effects such as fatigue because there are possible indicators for fatigued workers in the dataset \cite{rabiger2018how}.

\subsection{Building Predictors}
\label{sec:building_predictors}
\begin{figure}[htb]
\centering
\includegraphics[width=.8\columnwidth]{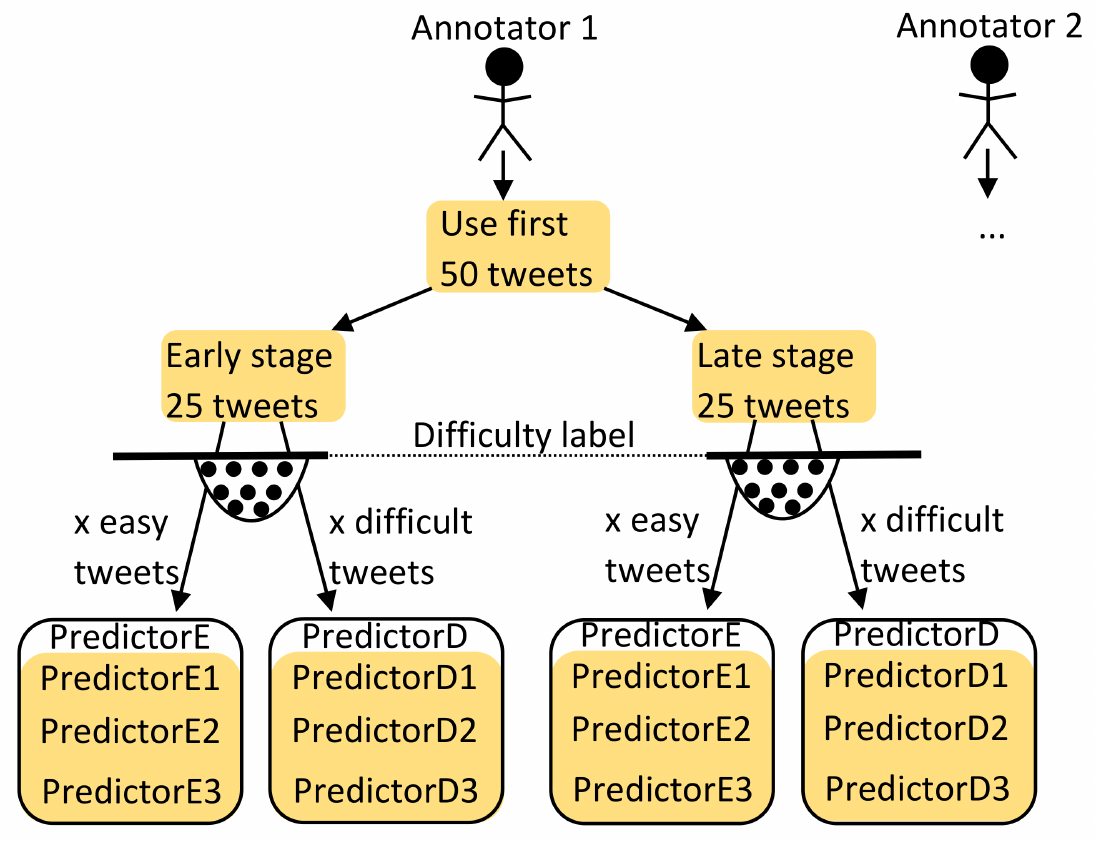}
\caption{Overview how predictors, using $x$ tweets for training, are built for a single crowd worker.}
\label{fig:schema2}
\end{figure}

One sentiment predictor (kNN) is trained per crowd worker in \MD{} and \SU{} because sentiment analysis is subjective.
The exact training procedure of \prednd{}and \predd{} for a single crowd worker is illustrated schematically in Figure~\ref{fig:schema2}. 
The training set (containing only \ambiguous{} or only \nonambiguous{} tweets) is derived from Tweets 1-25 in the \learningphase{} and once from tweets 26-50 in the \exploitationphase. 
This leads effectively to four datasets per worker to which we refer in the remainder as strata, namely:
\begin{enumerate}
\item EARLY\_EASY: \nonambiguous{} tweets that were labeled in a worker's \learningphase
\item EARLY\_DIFFICULT: \ambiguous{} tweets that were labeled in a worker's \learningphase
\item LATE\_EASY: \nonambiguous{} tweets that were labeled in a worker's \exploitationphase
\item LATE\_DIFFICULT: \ambiguous{} tweets that were labeled in a worker's \exploitationphase
\end{enumerate}
Hierarchical learning is performed by training in total six predictors (two predictors are trained per hierarchy level). Note that we introduced an extra label besides the sentiment labels to indicate that no label exists on a certain hierarchy level. This is necessary as \irrelevant{} tweets have only a label on the top-most hierarchy level. To assess the performance of the trained predictors in terms of hierarchical F1-scores (micro-averaged over all workers in a stratum), the labels of the remaining tweets in a worker's stratum are estimated per hierarchy level. For example, if \prednd{} is trained on five tweets that an annotator labeled in EARLY\_EASY, it will be evaluated on her remaining 20 labeled tweets.

\subsection{Testing the Meaningfulness of Observed Patterns}
\label{sec:testing_meaningfulness}
Since we vary many parameters in our simulation, it will be hard to depict all plotted configurations. Instead, our main goal is to identify patterns that hold over different configurations as these are more likely to be meaningful. 
We will report all our  results in an encoded form to make finding patterns more straightforward. Instead of showing how the F1-scores of the predictors develop when varying the size of the training set, we simply state if one of the two resulting F1-curves dominates the other one. In that case there are three possible outcomes: either curve dominates the other one or there is a tie.  The details about the encoding are explained in Section~\ref{sec:rq1}. However, reporting these encoded results permits us to test if there are significant differences in the proportions of the three outcomes using the two-tailed Fisher's exact test. Fisher's exact test (instead of a chi-square test) is suitable since some of the outcomes occur rarely.

\section{Results}
First, we show some sample F1-curves of the trained predictors because afterwards we encode them into a compressed form to be able to report all of our results. This allows to identify certain trends whose statistical significance we examine thereafter.

\subsection{Observed Patterns in the Simulation Experiment}
\label{sec:rq1}
This section addresses \rejectedrqone.
In our dataset, \nonambiguous{} and \ambiguous{} tweets are roughly equally distributed, with \nonambiguous{} tweets (according to Eq.~\ref{eq:ds}) accounting for 50\% to 57\% of the tweets depending on the stratum as illustrated in Table~\ref{tab:labels}. That means the classes are sufficiently balanced, thus there is no need to take any special countermeasures in the classification task. 

\begin{table}[htb]
\centering
\begin{subtable}{\columnwidth}
\centering
    \begin{tabular}[bth]{ccc} \toprule
     & \MD & \SU \\\midrule
    Easy & 68 (50.4\%) & 93 (57.4\%)\\
    Difficult & 67  (49.6\%) & 69 (42.6\%)\\\bottomrule
    \end{tabular}
    \caption*{Early stage}
   \end{subtable}
    \par\medskip 
    \begin{subtable}{\columnwidth}
    \centering
    \begin{tabular}[bth]{ccc} \toprule
     & \MD & \SU \\\midrule
    Easy & 78 (55.3\%) & 86 (54.3\%)\\
    Difficult & 63  (44.7\%) & 72 (45.7\%)\\\bottomrule
    \end{tabular}
    \caption*{Late stage}
    \end{subtable}
    \caption{Absolute numbers and percentages of \nonambiguous/\ambiguous{} tweets per stratum for both groups, \MD{} and \SU.}
\label{tab:labels}
\end{table}

First, we show some sample F1-curves of the trained predictors because afterwards we encode them into a compressed form to be able to report all of our results. This allows to identify certain trends whose statistical significance we examine thereafter.

\begin{figure}[htb]
    \begin{subfigure}[t]{1\columnwidth}
        \centering
        \includegraphics[width=1.0\columnwidth]{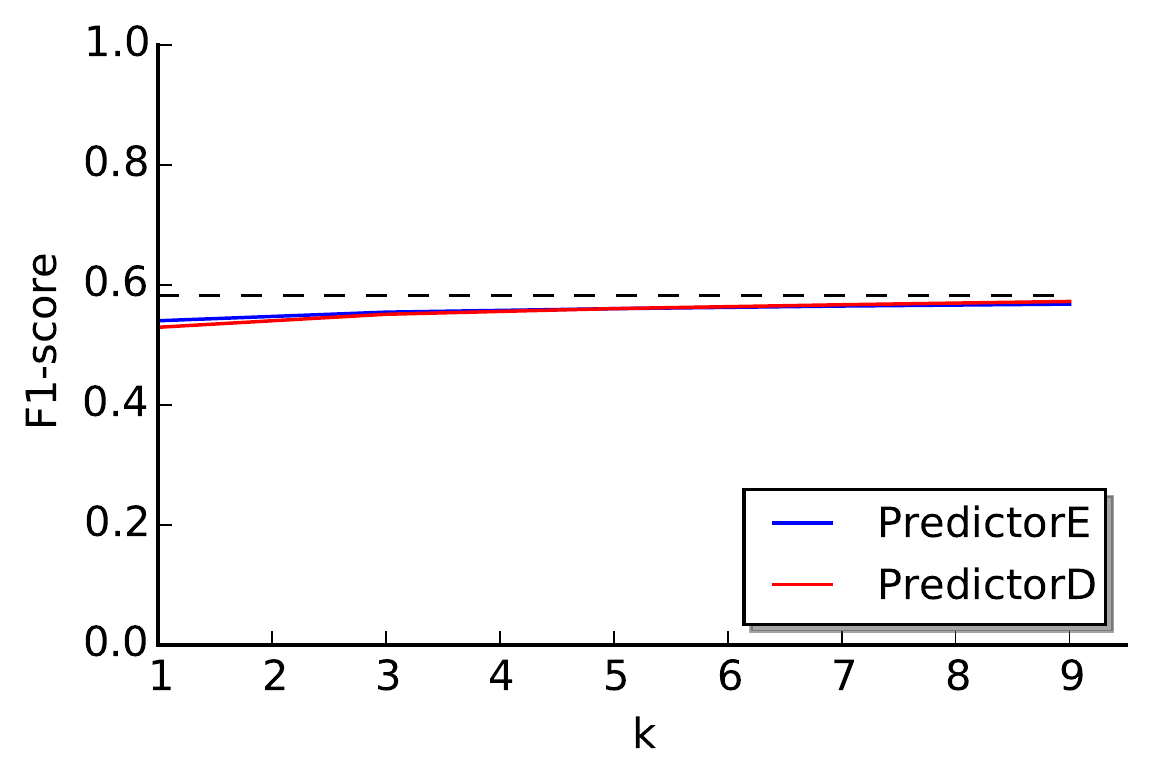}
        \caption{\MD}
    \end{subfigure}
    \begin{subfigure}[t]{1\columnwidth}
        \centering
        \includegraphics[width=1.0\columnwidth]{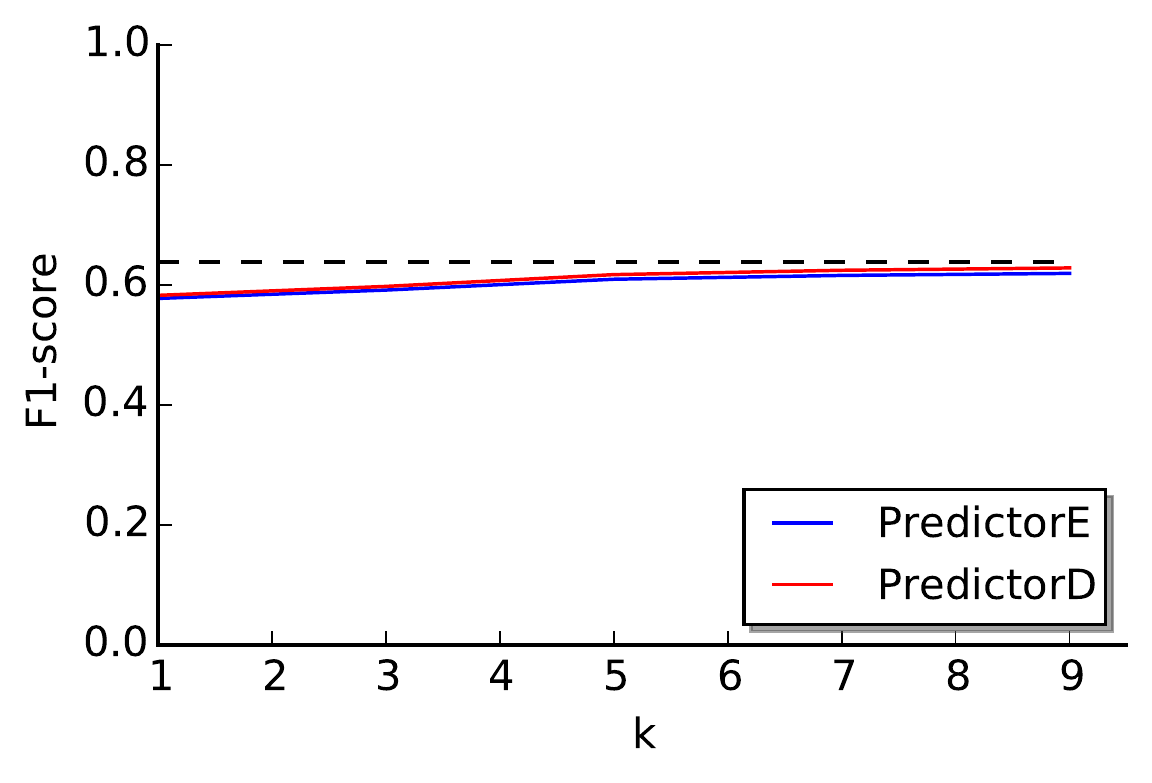}
        \caption{\SU}
    \end{subfigure}
    \caption{F1-scores of kNN with varying k. For each annotator the training set comprises eight (\nonambiguous/\ambiguous) tweets of the early phase.}
    \label{fig:early_stage}
\end{figure}

\begin{figure}[htb]
    \centering
     \begin{subfigure}[t]{1\columnwidth}
        \centering
        \includegraphics[width=1.0\columnwidth]{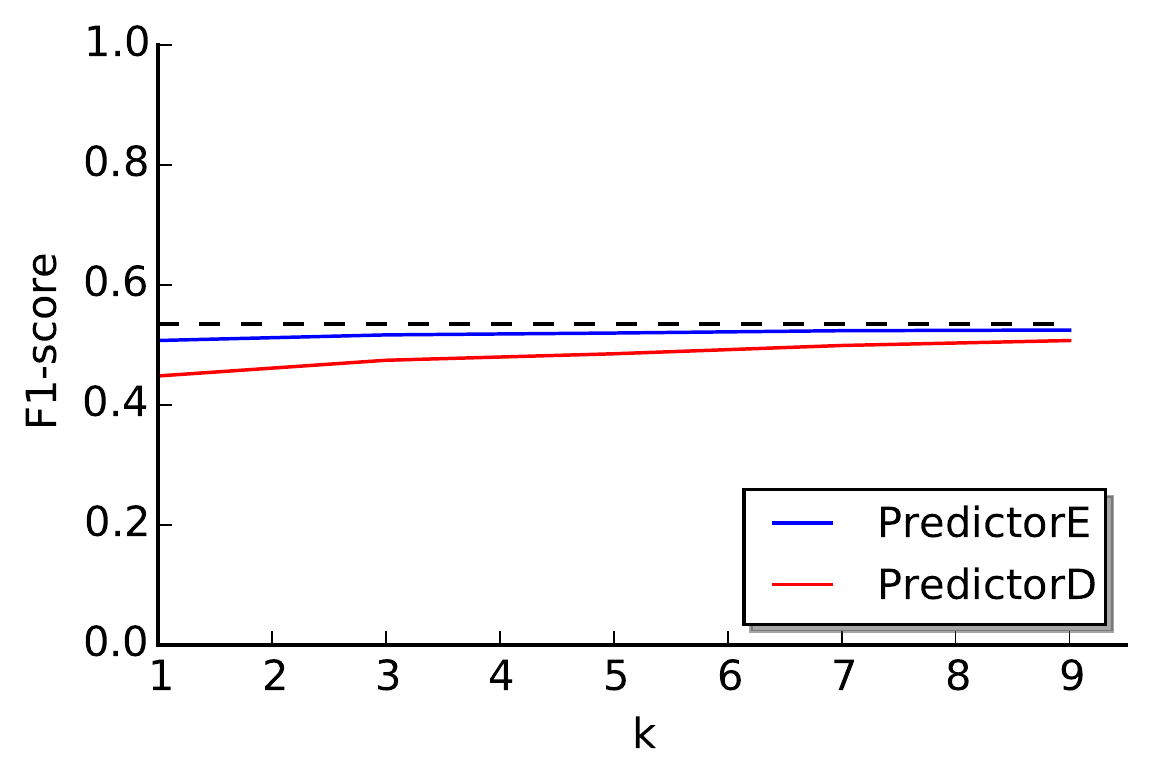}
        \caption{\MD}
    \end{subfigure}
    \begin{subfigure}[t]{1.0\columnwidth}
        \centering
        \includegraphics[width=1\columnwidth]{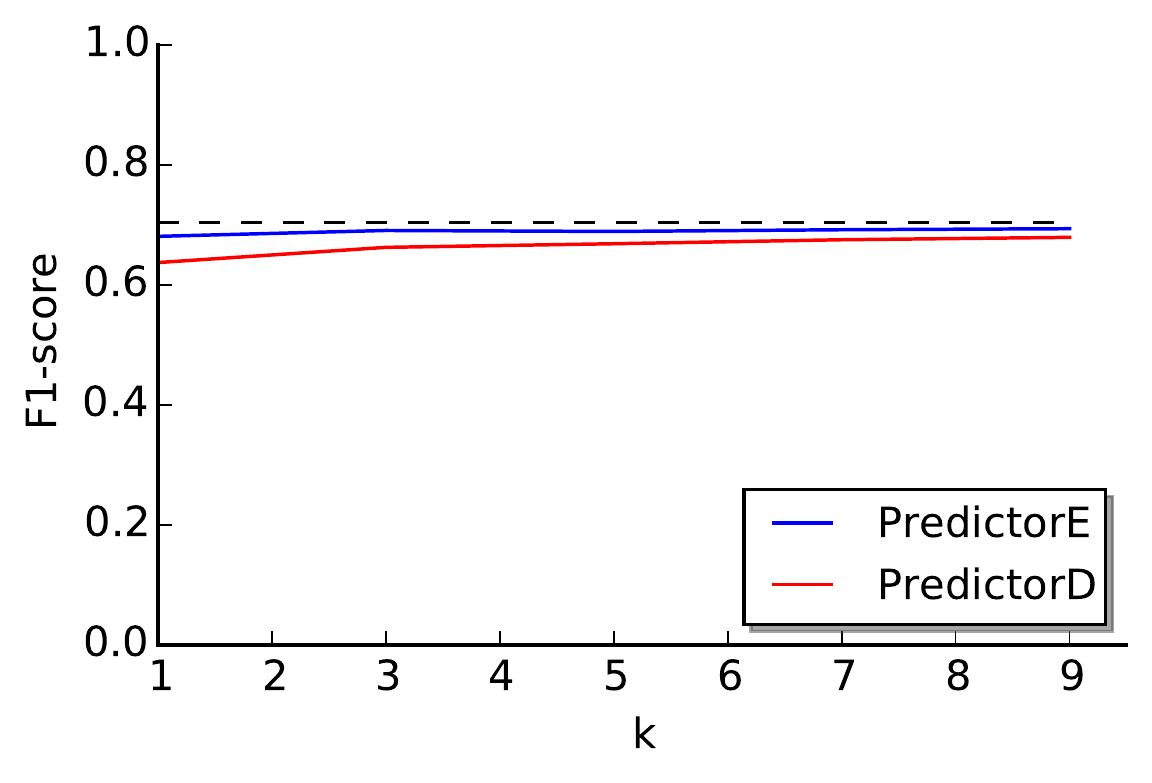}
        \caption{\SU}
    \end{subfigure}
    \caption{F1-scores of kNN with varying $k$. For each annotator the training set comprises eight (\nonambiguous/\ambiguous) tweets of the late phase.}
    \label{fig:late_stage}
\end{figure}

We show the F1-curves of the kNN predictors trained on eight tweets per worker for the four strata while varying $k$, the number of neighbors in kNN. The predictors utilize edit distance as a similarity metric. In Figure~\ref{fig:early_stage}, the F1-curves of \prednd{} trained on EARLY\_EASY and \predd{} trained on EARLY\_DIFFICULT are shown for \MD{} and \SU{}. In that case both predictors perform equally well. This observation holds in both groups and will be encoded as (T)ie in the compressed form. We note that the differences between the F1-curves in the \learningphase{} are generally small.
The corresponding F1-scores for the \exploitationphase{} of \MD{} and \SU{} are depicted in Figure~\ref{fig:late_stage} using the same setup as described before. This means that now the performances of \prednd{} trained on LATE\_EASY and \predd{} trained on LATE\_DIFFICULT are evaluated. This time, \prednd{} outperforms \predd. This behavior is consistent in \MD{} and \SU{} and will be encoded as (E)asy in the compressed representation. In this specific case, the F1-scores of \prednd{} in \SU{} are between 1.5\% and 4.5\% higher than in \predd. In \MD{}, \prednd{} achieves between 2\% and 6\% better F1-scores than \predd. We note that the differences between the F1-curves tend to be larger if \prednd{} outperforms \predd. If \predd{} wins, both F1-curves are close to each other. 
In both figures it seems that considering more neighbors for predictions mainly improves the F1-scores of \predd{} but not \prednd. This could indicate that less workers are necessary to label \nonambiguous{} tweets as opposed to \ambiguous{} ones. 

We report the outcomes of the remaining F1-curves of the predictors for the four strata with varying training sets containing between two and ten tweets as follows. At all times we compare in a stratum the F1-scores of \prednd{} and \predd{} while varying $k$. We encode each outcome as follows (abbreviation in parentheses):
\begin{itemize}
    \item (T)ie (both predictors exhibit the same F1-scores), 
    \item (E)asy (\prednd{} outperforms \predd), 
    \item (D)idifficult (\predd{} outperforms \prednd).
\end{itemize}

Each table contains the encoded outcomes over training sets comprising between two and ten tweets using different distance metrics. More specifically, Table~\ref{tab:edit} depicts the outcomes for the edit distance, Table~\ref{tab:subsequence} shows the outcomes for the longest common subsequence, and Table~\ref{tab:substring} gives the results for the longest common substring. 
One tendency in these tables is that the likelihood of seeing T drops as the number of tweets used for learning increases. We suspect that this
phenomenon occurs because a small number of training tweets leads to a
poor predictor performance anyway, no matter whether these tweets were
\nonambiguous{} or \ambiguous. As soon as the number of training tweets increases,
the difference becomes apparent, whereupon it becomes more likely that \prednd{} is the best one.

We juxtaposed the winner predictors between the two groups \MD{} and \SU{}
once for the \learningphase{} and once for the \exploitationphase. The numbers are too
small to deliver robust results, but we observe a general tendency:
\prednd{} is more often the winner in the \exploitationphase{} for \SU{} than for
\MD{}. This could be seen as an indication that \SU{} learned
faster, but the phenomenon can also be explained by differences in size
between the two groups: \MD{} is smaller and thus more vulnerable
to variations in the performance of the individual annotators.
Another related pattern across all groups is that T occurs frequently in the \learningphase, while E tends to appear more often in the \exploitationphase.

\begin{table}[htb]
    \begin{subtable}{.48\columnwidth}
        \centering
        
        \begin{tabular}[bth]{cccccccccc} \toprule
         & 2 & 3 & 4 & 5 & 6 & 7 & 8 & 9 & 10 \\\midrule
        Early & T & T & T & D & D & D & E & E & E\\
        Late & T & T & T & E & E & E & E & E & E\\\bottomrule
        \end{tabular}
        \caption*{\MD}
    \end{subtable}
    \par\medskip 
    \begin{subtable}{.48\columnwidth}
        \centering
        \begin{tabular}[bth]{cccccccccc} \toprule
         & 2 & 3 & 4 & 5 & 6 & 7 & 8 & 9 & 10 \\\midrule
        Early & T & T & T & T & T & D & D & T & E\\
        Late & T & E & E & E & E & E & E & E & E\\\bottomrule
        \end{tabular}
        \caption*{\SU}
    \end{subtable}
\caption{Outcomes for the different strata using kNN with edit distance and a varying number of tweets (2-10) in the training set of each annotator.}
\label{tab:edit}
\end{table}

\begin{table}[htb]
    \begin{subtable}{.48\columnwidth}
        \centering
        \begin{tabular}[bth]{cccccccccc} \toprule
         & 2 & 3 & 4 & 5 & 6 & 7 & 8 & 9 & 10 \\\midrule
         Early & T & T & T & D & T & T & E & E & E\\
         Late & T & D & T & T & T & E & E & E & E\\\bottomrule
        \end{tabular}
        \caption*{\MD}
    \end{subtable}
    \par\medskip 
    \begin{subtable}{.48\columnwidth}
        \centering
        \begin{tabular}[bth]{cccccccccc} \toprule
         & 2 & 3 & 4 & 5 & 6 & 7 & 8 & 9 & 10 \\\midrule
        Early & T & T & T & T & T & D & D & T & E\\
        Late & T & E & E & D & E & D & E & E & E\\\bottomrule
        \end{tabular}
        \caption*{\SU}
    \end{subtable}
\caption{Outcomes for the different strata using kNN with longest common subsequence and a varying number (2-10) of tweets in the training set of each annotator.}
\label{tab:subsequence}
\end{table}

\begin{table}[htb]

    \begin{subtable}{.48\columnwidth}
        \centering
        \begin{tabular}[bth]{cccccccccc} \toprule
         & 2 & 3 & 4 & 5 & 6 & 7 & 8 & 9 & 10 \\\midrule
         Early & T & T  & T & T & T & E & E & E & E\\
         Late & T & D & T & E & E & T & E & E & E\\\bottomrule
        \end{tabular}
        \caption*{\MD}
    \end{subtable}
    \par\medskip 
    \begin{subtable}{.48\columnwidth}
        \centering
        \begin{tabular}[bth]{cccccccccc} \toprule
         & 2 & 3 & 4 & 5 & 6 & 7 & 8 & 9 & 10 \\\midrule
        Early & T & T & T & T & T & D & D & T & E\\
        Late & T & E & E & E & E & D & E & E & E\\\bottomrule
        \end{tabular}
        \caption*{\SU}
    \end{subtable}
\caption{Outcomes for the different strata using kNN with longest common substring and a varying number of tweets (2-10) in the training set of each annotator.}
\label{tab:substring}
\end{table}

\subsection{Significance of Observed Patterns}
\label{sec:simulation_patterns_significance}
To analyze the meaningfulness of these patterns according to \rejectedrqtwo, we run the two-tailed Fisher's exact test to see if the differences in the proportions of the outcomes are significant as described in Section~\ref{sec:testing_meaningfulness}. For comparing all pairwise proportions, our null hypotheses to be tested are: there is no difference in the proportion of E and D (T and E) (T and D) between \learningphase{} and \exploitationphase. The proportions are displayed in Table~\ref{tab:proportions} and were obtained by adding up the outcomes from Tables~\ref{tab:edit}-\ref{tab:substring}. Using $\alpha=0.05$ as significance level, we obtain the following results. 

\begin{table}[htb]
    \begin{subtable}{.48\columnwidth}
        \centering
        \begin{tabular}[bth]{ccc} \toprule
         & Early & Late\\\midrule
        T & 31 & 12 \\
        E & 13 & 36 \\\bottomrule
        \end{tabular}
        \caption*{E vs. T}
    \end{subtable}
    \begin{subtable}{.48\columnwidth}
        \centering
    \begin{tabular}[bth]{ccc} \toprule
         & Early & Late\\\midrule
        E & 13 & 36 \\
        D & 10 & 6 \\\bottomrule
        \end{tabular}
        \caption*{E vs. D}
    \end{subtable}
    \par\medskip 
    \begin{subtable}{.48\columnwidth}
        \centering
        \begin{tabular}[bth]{ccc} \toprule
         & Early & Late\\\midrule
        T & 31 & 12 \\
        D & 10 & 6 \\\bottomrule
        \end{tabular}
        \caption*{T vs. D}
    \end{subtable}
\caption{Occurrences of the encoded outcomes in an annotator's early and late phase.}
\label{tab:proportions}
\end{table}
The proportions of E and T are significantly different in the early and late phase ($p<0.0001$). This suggests that ties between predictors occur more frequently in the early phase, while \prednd{} outperforms \predd{} significantly more often in the late phase.
Likewise, the proportions of E and D differ significantly ($p<0.02$) across both phases, which means that neither of \prednd{} nor \predd{} wins significantly more frequently in the early phase, while in the later phase \prednd{} outperforms \predd{} significantly more often. When it comes to the proportions of T and D, no significant differences exist in the proportions ($p>0.5$). 
Thus, the significance tests confirm our intuition about the existing patterns in the results, namely that T occurs mainly in the early phase, E in the late phase and D appears rarely in both phases.

\section{Discussion}
\label{sec:discussion}
The results of our preliminary study suggest that there is indeed a connection between the difficulty of tweets and the reliability of the labels that annotators assigned to them. More specifically, the label reliability of \nonambiguous{} tweets seems higher, because predictors trained on them achieve higher F1-scores. However, this holds only for an annotator's \exploitationphase, i.e. after annotators have already labeled 25 other tweets. In the \learningphase, i.e. for the first 25 tweets, our results do not show any evidence for such a relationship. 
One possible explanation for this result could be that the labels workers assign in their \learningphase{} \cite{settles2008active,zhu2010analysis,gadiraju2015training,maddalena2016crowdsourcing,rabiger2018how} are generally of lower quality during that period \cite{maddalena2016crowdsourcing, rabiger2018how}. Therefore, the higher level of noisy, low-quality labels in the \learningphase{} could be masking the effect of tweet difficulty on label reliability in an annotator's \learningphase. 

It would be interesting to examine this hypothesis using a slightly different experiment setup than our current one in a new study: first, workers complete a labeling task in their first annotation session (same setup as in \cite{rabiger2018how}) and after a short break, they repeat the task with new tweets in a second session. If the noisy, low-quality labels due to the \learningphase{} masked the relationship between tweet difficulty and label reliability in the\learningphase{} of the first session, in the second session we would expect to see a pattern similar to the one we reported for the \exploitationphase{} in this thesis, because workers should not have to go through another \learningphase, assuming the break between two sessions is not too long. 
However, given that crowd workers tend to complete many micro-tasks, they will quickly reach their \exploitationphase, meaning that labeling easier tweets will increase the reliability of assigned labels in practice. 

This motivates the idea of devising a tweet difficulty predictor to estimate the difficulty of unknown tweets for which a host of applications exist \cite{rabiger2018predicting}. 
We plan to apply this predictor as a filter before an actual crowdsourcing task. Given a large dataset, one could crowdsource a small seed first to train the difficulty predictor. It then estimates the level of difficulty in the unlabeled dataset and only tweets which are estimated to be easy would be crowdsourced.  This could also complement the approach proposed by Whitehill et al. \cite{whitehill2009whose} for aggregating crowdsourced labels more accurately because the prior for document difficulty in their probabilistic method could be tweaked such that \nonambiguous{} documents are more likely to occur in the dataset. 
Building such a difficulty predictor on a small seed set would also benefit active learning techniques, as they could be invoked only on easy tweets to obtain reliable labels from experts. Here the difficulty predictor would be used before invoking an active learning algorithm only for easy tweets. Furthermore, incorporating tweet difficulty into cost models in active learning, that estimate the costs for acquiring labels for unlabeled tweets, could enhance the models' accuracy.

Reducing the dataset size by filtering out \ambiguous{} tweets could potentially increase the retention rate of the crowdsourcing task as workers might become less frustrated since micro-tasks can be completed with more ease. Furthermore, crowdsourcing a smaller dataset could save budget that will not be spent on \ambiguous{} tweets. 
Even more budget could be saved if less crowd workers would be allocated to \nonambiguous{} tweets, similar to \cite{sameki2016dynamic}. 
Another way of using such a tweet difficulty predictor would be to assign \nonambiguous{} tweets for labeling to inexperienced workers and \ambiguous{} ones to experts \cite{kolobov2013joint}. The associated monetary compensation could possibly also vary depending on the level of expertise of crowd workers. This is related to the problem of optimal task routing in crowdsourcing where suitable workers should be identified for micro-tasks. For example, in \cite{goncalves2017task} workers' cognitive abilities are used to match them to suitable tasks. This works for language fluency and visual tasks, but has not been tested for other types of tasks, such as sentiment analysis. If tweets are involved, a tweet difficulty predictor could complement this approach.

We note several limitations in our preliminary study. First, our dataset was relatively small. Nevertheless, the tweets we used were diverse and we performed our experiment independently in two different locations.
Second, we investigated a single labeling task and it could bias the results. For example, in other tasks \nonambiguous{} tweets might not be diverse enough to train good predictors. However, if sufficiently diverse tweets exist for a labeling task, we believe that our results will hold. Third, we evaluated only one predictor, kNN. Thus, replicating this experiment on a larger scale with more diverse predictors would help establish our findings. Our dataset\footnote{\url{https://www.researchgate.net/publication/325180810_Infsci2017_dataset}} and source code\footnote{\url{https://github.com/fensta/PrelimStudy}} are publicly available.

\section{Conclusion}
In this preliminary study we examined how tweet difficulty affects the reliability of labels that annotators assign. The experiment we designed to investigate this hypothesis was performed independently in two locations and we obtained consistent empirical results. They suggest that the labels assigned to \nonambiguous{} tweets are more reliable, but only if the annotators are familiar with the labeling task, i.e. they had labeled a certain number of tweets before. This observation implies that the performance of predictors could be theoretically enhanced by devising a predictor that can estimate the difficulty of tweets in advance. 
Due to its benefits for crowdsourcing and active learning, we plan to develop a method that employs such a tweet difficulty predictor at its core in the future \cite{rabiger2018predicting}.
Another subject for future investigation is the question of diversity in \nonambiguous{} tweets: do the \nonambiguous{} tweets in a labeling task always suffice to train meaningful predictors?


\begin{thebibliography}{10}

\bibitem{alonso2015predicting}
H{\'e}ctor~Mart{\i}nez Alonso, Anders Johannsen, Oier~Lopez de~Lacalle, and
  Eneko Agirre.
\newblock Predicting word sense annotation agreement.
\newblock In {\em Workshop on Linking Models of Lexical, Sentential and
  Discourse-level Semantics (LSDSem)}, page~89, 2015.

\bibitem{alonso2009can}
Omar Alonso and Stefano Mizzaro.
\newblock Can we get rid of trec assessors? using mechanical turk for relevance
  assessment.
\newblock In {\em Proceedings of the SIGIR 2009 Workshop on the Future of IR
  Evaluation}, volume~15, page~16, 2009.

\bibitem{bontcheva2017crowdsourcing}
Kalina Bontcheva, Leon Derczynski, and Ian Roberts.
\newblock Crowdsourcing named entity recognition and entity linking corpora.
\newblock In {\em Handbook of Linguistic Annotation}, pages 875--892. Springer,
  2017.

\bibitem{carmel2006makes}
David Carmel, Elad Yom-Tov, Adam Darlow, and Dan Pelleg.
\newblock What makes a query difficult?
\newblock In {\em Proceedings of the 29th annual international ACM SIGIR
  conference on Research and development in information retrieval}, pages
  390--397. ACM, 2006.

\bibitem{culotta2005reducing}
Aron Culotta and Andrew McCallum.
\newblock Reducing labeling effort for structured prediction tasks.
\newblock In {\em AAAI}, volume~5, pages 746--751, 2005.

\bibitem{gadiraju2015training}
Ujwal Gadiraju, Besnik Fetahu, and Ricardo Kawase.
\newblock Training workers for improving performance in crowdsourcing
  microtasks.
\newblock In {\em Design for Teaching and Learning in a Networked World}, pages
  100--114. Springer, 2015.

\bibitem{gadiraju2014taxonomy}
Ujwal Gadiraju, Ricardo Kawase, and Stefan Dietze.
\newblock A taxonomy of microtasks on the web.
\newblock In {\em Proceedings of the 25th ACM conference on Hypertext and
  social media}, pages 218--223. ACM, 2014.

\bibitem{gan2017incentivize}
Xiaoying Gan, Xiong Wang, Wenhao Niu, Gai Hang, Xiaohua Tian, Xinbing Wang, and
  Jun Xu.
\newblock Incentivize multi-class crowd labeling under budget constraint.
\newblock {\em IEEE Journal on Selected Areas in Communications},
  35(4):893--905, 2017.

\bibitem{goncalves2017task}
Jorge Goncalves, Michael Feldman, Subingqian Hu, Vassilis Kostakos, and Abraham
  Bernstein.
\newblock Task routing and assignment in crowdsourcing based on cognitive
  abilities.
\newblock In {\em Proceedings of the 26th International Conference on World
  Wide Web Companion}, pages 1023--1031. International World Wide Web
  Conferences Steering Committee, 2017.

\bibitem{kazai2013analysis}
Gabriella Kazai, Jaap Kamps, and Natasa Milic-Frayling.
\newblock An analysis of human factors and label accuracy in crowdsourcing
  relevance judgments.
\newblock {\em Information retrieval}, 16(2):138--178, 2013.

\bibitem{kiritchenko2006learning}
Svetlana Kiritchenko, Stan Matwin, Richard Nock, and A~Fazel Famili.
\newblock Learning and evaluation in the presence of class hierarchies:
  Application to text categorization.
\newblock In {\em Canadian Conference on AI}, volume 2006, pages 395--406.
  Springer, 2006.

\bibitem{kolobov2013joint}
Andrey Kolobov, Daniel~S Weld, et~al.
\newblock Joint crowdsourcing of multiple tasks.
\newblock In {\em First AAAI Conference on Human Computation and
  Crowdsourcing}, pages 36--37, 2013.

\bibitem{maddalena2016crowdsourcing}
Eddy Maddalena, Marco Basaldella, Dario De~Nart, Dante Degl'Innocenti, Stefano
  Mizzaro, and Gianluca Demartini.
\newblock Crowdsourcing relevance assessments: The unexpected benefits of
  limiting the time to judge.
\newblock In {\em Fourth AAAI Conference on Human Computation and
  Crowdsourcing}, 2016.

\bibitem{martinez2013document}
Miguel Martinez-Alvarez, Alejandro Bellogin, and Thomas Roelleke.
\newblock Document difficulty framework for semi-automatic text classification.
\newblock In {\em International Conference on Data Warehousing and Knowledge
  Discovery}, pages 110--121. Springer, 2013.

\bibitem{paukkeri2013assessing}
Mari-Sanna Paukkeri, Marja Ollikainen, and Timo Honkela.
\newblock Assessing user-specific difficulty of documents.
\newblock {\em Information Processing \& Management}, 49(1):198--212, 2013.

\bibitem{rabiger2018predicting}
Stefan R{\"a}biger, Gizem Gezici, Myra Spliliopoulou, and Y\"{u}cel Sayg{\i}n.
\newblock Predicting worker disagreement for more effective crowd labeling.
\newblock In {\em 2018 IEEE 5th International Conference on Data Science and
  Advanced Analytics (DSAA)}. IEEE, 2018.

\bibitem{rabiger2018how}
Stefan R{\"a}biger, Myra Spiliopoulou, and Y\"{u}cel Sayg{\i}n.
\newblock How do annotators label short texts? toward understanding the
  temporal dynamics of tweet labeling.
\newblock {\em Information Sciences}, 457-458:29 -- 47, 2018.

\bibitem{salomoni2015crowdsourcing}
Paola Salomoni, Catia Prandi, Marco Roccetti, Valentina Nisi, and N~Jardim
  Nunes.
\newblock Crowdsourcing urban accessibility:: Some preliminary experiences with
  results.
\newblock In {\em Proceedings of the 11th Biannual Conference on Italian SIGCHI
  Chapter}, pages 130--133. ACM, 2015.

\bibitem{sameki2016dynamic}
Mehrnoosh Sameki, Mattia Gentil, Kate~K Mays, Lei Guo, and Margrit Betke.
\newblock Dynamic allocation of crowd contributions for sentiment analysis
  during the 2016 us presidential election.
\newblock {\em arXiv preprint arXiv:1608.08953}, 2016.

\bibitem{settles2012active}
Burr Settles.
\newblock Active learning.
\newblock {\em Synthesis Lectures on Artificial Intelligence and Machine
  Learning}, 6(1):1--114, 2012.

\bibitem{settles2008active}
Burr Settles, Mark Craven, and Lewis Friedland.
\newblock Active learning with real annotation costs.
\newblock In {\em Proceedings of the NIPS workshop on cost-sensitive learning},
  pages 1--10, 2008.

\bibitem{whitehill2009whose}
Jacob Whitehill, Ting fan Wu, Jacob Bergsma, Javier~R. Movellan, and Paul~L.
  Ruvolo.
\newblock Whose vote should count more: Optimal integration of labels from
  labelers of unknown expertise.
\newblock In Y.~Bengio, D.~Schuurmans, J.~D. Lafferty, C.~K.~I. Williams, and
  A.~Culotta, editors, {\em Advances in Neural Information Processing Systems
  22}, pages 2035--2043. Curran Associates, Inc., 2009.

\bibitem{yang2013freeloc}
Sungwon Yang, Pralav Dessai, Mansi Verma, and Mario Gerla.
\newblock Freeloc: Calibration-free crowdsourced indoor localization.
\newblock In {\em INFOCOM, 2013 Proceedings IEEE}, pages 2481--2489. IEEE,
  2013.

\bibitem{yang2016enhancing}
Xiaoyan Yang, Shanshan Ying, Wenzhe Yu, Rong Zhang, and Zhenjie Zhang.
\newblock Enhancing topic modeling on short texts with crowdsourcing.
\newblock In {\em Asian Conference on Machine Learning}, pages 33--48, 2016.

\bibitem{zhu2010analysis}
Dongqing Zhu and Ben Carterette.
\newblock An analysis of assessor behavior in crowdsourced preference
  judgments.
\newblock In {\em SIGIR 2010 workshop on crowdsourcing for search evaluation},
  pages 17--20, 2010.

\end{thebibliography}
\end{document}